\documentclass{elsart}
\usepackage{amsmath}
\usepackage{amsfonts,amssymb}
\newsymbol\lessim 132E
\usepackage{graphicx}
\usepackage{times}
\begin{document}

\begin{frontmatter}

\title{Linear transmitter with correlated noises}

\author{P.  F.  G\'ora}
\ead{gora@if.uj.edu.pl}
\address{M.~Smoluchowski Institute of Physics and Complex Systems Research Center, 
Jagellonian University, Reymonta~4, 30--059~Krak\'ow, Poland}

\begin{abstract}
A linear transmitter with correlated Gaussian white additive and multiplicative
noises and a periodic signal coupled either additively or multiplicatively
is considered. The correlations have a weak destructive effect in case of the
additive signal and a strong constructive effect in case of the multiplicative
one. Analytical results for the linear transmitter with the additive signal
agree with those obtained previously by a different approach.
We also show how to analytically calculate certain expectation values
involving exponentials of Gaussian processes.
\end{abstract}

\begin{keyword}
Correlated Gaussian noises \sep stochastic resonance \sep signal-to-noise ratio
\PACS 05.40.Ca
\end{keyword}

\end{frontmatter}

\section{Introduction}

The fact that noise can play a constructive role in many physical systems
is now widely recognized. The best-known examples of such phenomena 
are the stochastic resonance (SR) \cite{SR}
and Brownian ratchets \cite{ratchet}. SR is a phenomenon in which the response
of a dynamical system is optimized by the presence of a specific level of noise
and has been detected in so many seemingly different systems that it has been
claimed to be ``an inherent property of rate-modulated series of events'' 
\cite{Bezrukov1}. However, it has been recently suggested that the functioning of 
important natural devices, e.g., communication and information
processing in neural systems or subthreshold signal detection
in biological receptors, rely on phase synchronization rather
than stochastic resonance \cite{Freund}. The SR is usually measured by
the signal-to-noise ratio (SNR) for the output signal but new
measures are being proposed and discussed \cite{Freund,Bezrukov2,gain}.

One of the issues that is now
debated and  that is an origin of many unexpected physical phenomena,
is the role of correlations between various noises present in the system. 
This point has been particularly stressed in research on Duffing oscillator
with additive and multiplicative noises \cite{Telejko,duffing}, where 
correlations are responsible for huge changes in the activation rate,
on Brownian ratchets \cite{Luczka}, where correlations between the
additive and multiplicative noises can lead to current reversal, and
on a nonlinear rotator \cite{singh}, where full correlations between two
noises amount to nullifying one of them, leaving one of the states
effectively noise-free and allowing for the passage of 
incoming signals of vanishing intensity. 
Other contexts in which correlations between
various noise terms appear to be important include a quantum dimer 
under the influence of colored noises \cite{oldPRE}, 
a coupled neuron network \cite{Bambi},
synchronization of chaotic oscillators \cite{Kiss}, detecting the 
gravitational background \cite{gravity}, and medical imaging \cite{imaging}.
The effect of correlations between a Gaussian white noise and a Gaussian
colored noise has been discussed in Ref.~\cite{Jia}.

It is sometimes very difficult to find analytical results for nonlinear
models and it is for that reason that models which admit rigorous solutions
are very interesting. One of such models is that of an overdamped linear 
transmitter, which in full form reads

\begin{equation}\label{chemphys:transmitter}
\dot a = (f_1(t) + f_2(t)\xi_1(t))a + f_3(t) + f_4(t)\xi_2(t)\,,
\end{equation}

\noindent where $\xi_{1,2}(t)$ are the noise terms.
It should be mentioned that while \eqref{chemphys:transmitter}
formally leads to a linear equation of motion with time-dependent 
coefficients, a multiplicative coupling between the noise and the 
system means a ``hidden'' nonlinearity: The noise is supposed to
represent many unobserved degrees of freedom coupled to
the transmitting process in a nonlinear manner.
Although chemical reaction with a fluctuating barrier provided the original
motivation for this model, with $a$ being the concentration of a reagent, 
the presence of the additive noise, representing the thermal bath acting on 
the system, and an additive deterministic signal possibly
lead to realizations that admit negative values of $a(t)$.
It, therefore, can no longer be regarded as the concentration of a chemical
species. Nevertheless, systems of this type can still describe processes of
a paramount physical interest like the electric potential across a molecular
membrane \cite{JSP} or provide a ``skeletal'' model of various enzymatic
reactions \cite{PRE2001}. This model, in less general forms, has also been 
discussed in Refs.~\cite{berdichevsky}. With one exception (see below),
possible implications of the presence of correlations between the two
noise terms in \eqref{chemphys:transmitter} have not been addressed so
far. Including such correlations is an obvious extension to the existing
research on models belonging to the general class \eqref{chemphys:transmitter}
and is the subject of the present paper.

There is one important exception to the general statement made in the 
preceding paragraph. Berdichevsky and Gitterman in Ref.~\cite{berdichevsky99} 
considered the linear transmitter with correlated noises and an additively
coupled signal. In order to calculate the correlation
function of the process $a(t)$ they obtained a whole hierarchy
$\left\langle a\right\rangle$, $\left\langle \xi a\right\rangle$,
$\left\langle \xi^2a\right\rangle$ etc, where $\xi$ is a noise term,
and needed a procedure to close the hierarchy.
They resorted to considering asymmetric, dichotomous, exponentially 
correlated noises and obtained results for Gaussian white noises only
as a limiting case. In this paper we show how to treat the Gaussian 
case directly; we think this approach is important and interesting, even
though the behavior predicted in Ref.~\cite{berdichevsky99} is certainly
more ``rich'' than that for a pure Gaussian white noise. We also slightly
differ with the authors of that reference in our analysis of final
results. In addition, we extend our previous discussion of a model
with a multiplicative signal \cite{PRE2001} to the case of correlated
noises. We find that correlations play an important role in this system.

This paper is organized as follows: In Section~\ref{chemphys:section-model} 
we introduce the model and discuss a couple of its formal features.
We give analytical results for the expectation value, the correlation function,
and the power spectrum of the system with an additive noise in 
Section~\ref{chemphys:section-additive}, and analytical and numerical results 
for the system with 
a~signal coupled to the system multiplicatively (parametrically) in
Section~\ref{chemphys:section-multiplicative}. A short discussion in 
Section~\ref{chemphys:section-discussion} follows.
Mathematical details are presented in the Appendix; although delegated
to an Appendix, these details represent an important part of this paper.

\section{The model}\label{chemphys:section-model}

Consider a system described by the following equation:

\begin{equation}\label{chemphys:model}
\dot a = - (k_a + \kappa\xi_1(t) + f(t))a + \lambda\xi_2(t) + g(t)\,.
\end{equation}

\noindent $\xi_1(t)$, $\xi_2(t)$ represent the multiplicative and additive noises,
respectively, acting on the system. $f(t)$ and $g(t)$ are deterministic external
stimulations; in the present paper we assume that \textit{either one} of them
vanishes identically. The constants $\kappa$ and $\lambda$ allow for an independent 
manipulation of the noises' strengths. The model \eqref{chemphys:model}
belongs to the general class of \eqref{chemphys:transmitter} but
is simplified to keep only its most important features.
The multiplicative noise represents
fluctuations of a barrier and the additive noise is a thermal bath acting on 
the system.

The noises acting on the system need not to be independent. On the contrary,
it is very likely that both the thermal bath and the process affecting the barrier 
height are correlated, as the molecular mechanisms responsible for these noises
are not independent. It is the objective of the present paper to examine the
effect of this correlation on the dynamics of the system. To keep the model as simple
as possible, we assume that the noises $\xi_1(t)$, $\xi_2(t)$ form a two dimensional
Gaussian process, point-correlated in time:

\begin{equation}\label{chemphys:correlations}
\left\langle
\left[\begin{array}{c}\xi_1(t)\\\xi_2(t)\end{array}\right]
\left[\begin{array}{cc}\xi_1(t^\prime)&\xi_2(t^\prime)\end{array}\right]
\right\rangle
=
\delta(t-t^\prime)
\left[\begin{array}{cc}
1&c\\c&1
\end{array}\right],
\end{equation}

\noindent where $c\in[-1,1]$ is the correlation coefficient. We now perform
the Cholesky decomposition \cite{golub} of the matrix in (\ref{chemphys:correlations}) 
and use it to construct
the noises from two independent processes:

\begin{equation}\label{chemphys:cholesky}
\left[\begin{array}{c}\xi_1(t)\\\xi_2(t)\end{array}\right]
=
\left[\begin{array}{cc}1&0\\c&\sqrt{1-c^2}\end{array}\right]
\left[\begin{array}{c}\xi(t)\\\eta(t)\end{array}\right].
\end{equation}

\noindent Here $\xi(t)$, $\eta(t)$ are two uncorrelated Gaussian white noises
(GWN): $\left\langle\xi(t)\right\rangle=0$, $\left\langle\eta(t)\right\rangle=0$,
$\left\langle\xi(t)\xi(t^\prime)\right\rangle=
\left\langle\eta(t)\eta(t^\prime)\right\rangle=\delta(t-t^\prime)$,
$\left\langle\xi(t)\eta(t^\prime)\right\rangle=0$, and all higher correlations
factorize. A more general mechanism of describing correlations between the
noises has been introduced in Ref.~\cite{Telejko}, but we will show that 
the dynamics of the system does not significantly depend on the detailed mechanism
of introducing the correlations.

We further assume that the external stimulations have a simple periodic form.
The evolution equation (\ref{chemphys:model}) thus becomes

\begin{equation}\label{chemphys:model2}
\dot a = -(k_a + \kappa\xi(t) + A_{\text{m}}\cos(\Omega t + \phi))a 
+ \lambda c\,\xi(t) + \lambda\sqrt{1-c^2}\,\eta(t) + A_{\text{a}}\cos(\Omega t + \phi),
\end{equation}

\noindent where $\phi$ is an initial phase of the signal. According to what
we have said above, only one of the amplitudes $A_{\text{m}}$, $A_{\text{a}}$
does not vanish.

The equation \eqref{chemphys:model2} can be formally solved. This solution can
be used to calculate its expectation value and the autocorrelation function.
Before we do that, we must first solve a couple of problems that will appear during
these calculations.

First, these calculations involve certain expectation values that, to our best knowledge,
have not been calculated so far: One has the form
$\left\langle \xi(t)e^{-\int_t^T p(t^\prime)\xi(t^\prime)dt^\prime}\right\rangle$
and the other
$\left\langle \xi(t_1)\xi(t_2)e^{-\int_t^T q(t^\prime)\xi(t^\prime)dt^\prime}\right\rangle$,
where $p(\cdot)$, $q(\cdot)$ are certain deterministic functions and $\xi(\cdot)$
is a GWN. We show how to calculate them in the Appendix.

Second, the correlation functions should be averaged over the initial phase of
the signals, as otherwise they will not be stationary \cite{JSP,key,Wiesenfeld}.
Therefore, we will use correlation functions of the form

\begin{equation}\label{chemphys:doubleaverage}
\left\langle\left\langle a(t_1)a(t_2)
\right\rangle\right\rangle =
\frac{1}{2\pi}\int\limits_0^{2\pi}\left\langle a(t_1)a(t_2)\right\rangle d\phi\,.
\end{equation}

\noindent The braces $\left\langle\cdots\right\rangle$ on the right-hand side
of Eq.~\eqref{chemphys:doubleaverage} stand for the average over realizations
of the noises.

Finally, calculating the power spectra of the processes described by
Eq.~\eqref{chemphys:model2} is our ultimate goal.
Let $a(t)$ be a solution to the equation \eqref{chemphys:model2}. 
We expect that it becomes stationary for $t\to\infty$. The simplest
estimate of the power spectrum is the square modulus of the Fourier 
transform of the process. If we were to
calculate the power spectrum numerically, we would calculate a particular
realization of $a(t)$, reject several (perhaps many) initial values
until the series becomes stationary, calculate the power spectrum of the
remaining series and finally average over the realization of the noises and 
the initial phase of the signal. We now follow this programme using the
analytical, not a numerical, solution. Thus, for the power spectrum we 
obtain

\begin{eqnarray}\label{chemphys:appB1}
P(\omega) &=& 
\lim\limits_{T\to\infty}\frac{1}{2T}
\lim\limits_{T_0\to\infty}
\left\langle\left\langle\left\vert\,
\int\limits_{T_0-T}^{T_0+T}
a(t)e^{i\omega t}dt
\right\vert^2
\right\rangle\right\rangle
\nonumber\\&=&
\lim\limits_{T\to\infty}\frac{1}{2T}
\lim\limits_{T_0\to\infty}
\int\limits_{T_0-T}^{T_0+T} dt_1
\int\limits_{T_0-T}^{T_0+T} dt_2
\left\langle\left\langle
a(t_1)a(t_2)
\right\rangle\right\rangle
e^{i\omega(t_1-t_2)}
\nonumber\\&=&
\lim\limits_{T\to\infty}\frac{1}{2T}
\lim\limits_{T_0\to\infty}
\int\limits_{-T}^T dt_1
\int\limits_{-T}^T dt_2
\left\langle\left\langle
a(T_0+t_1)a(T_0+t_2)
\right\rangle\right\rangle
\cos\omega(t_1-t_2)
.\nonumber\\
\end{eqnarray}

\noindent Note that in finite times approximants of the right-hand side
of Eq.~\eqref{chemphys:appB1}
we must always have $T_0\gg T\gg0$ and the limit $T_0\to\infty$ must be
taken first. After a simple change of variables, we obtain

\begin{eqnarray}\label{chemphys:appB2}
P(\omega)&=&
\lim\limits_{T\to\infty} \frac{1}{2T}
\int\limits_0^{2T} d\tau\,\cos\omega\tau \times{}
\nonumber\\&&
\lim\limits_{T_0\to\infty}
\int\limits_{\tau-2T}^{\tau+2T} dt
\left\langle\left\langle
a\left(T_0+{\textstyle\frac{1}{2}}t-{\textstyle\frac{1}{2}}\tau\right)
a\left(T_0+{\textstyle\frac{1}{2}}t+{\textstyle\frac{1}{2}}\tau\right)
\right\rangle\right\rangle.
\end{eqnarray}

\noindent If the process $a(t)$ becomes stationary in the asymptotic regime,
its autocorrelation function depends only on the difference of its arguments:

\begin{equation}\label{chemphys:appB3}
\lim\limits_{t\to\infty}
\left\langle\left\langle
a\left(t-{\textstyle\frac{1}{2}}\tau\right)
a\left(t+{\textstyle\frac{1}{2}}\tau\right)
\right\rangle\right\rangle
=C(\tau).
\end{equation}

\noindent We thus have

\begin{equation}\label{chemphys:appB4}
P(\omega) = 2\int\limits_0^\infty \cos\omega\tau\,C(\tau)\,d\tau\,.
\end{equation}

If the process $a(t)$ is stationary, we immediately have 

\begin{equation}\label{chemphys:lekcja}
C(\tau)=
\lim\limits_{t\to\infty}
\left\langle\left\langle
a\left(t-\tau\right)
a\left(t\right)
\right\rangle\right\rangle
=\lim\limits_{t\to\infty}
\left\langle\left\langle
a\left(t\right)
a\left(t+\tau\right)
\right\rangle\right\rangle
\end{equation}

\noindent and we can see that Eq.~\eqref{chemphys:appB4} is in the standard
Wiener-Khinchin form.

\section{An additive signal}\label{chemphys:section-additive}

We now turn to discussing specific forms of the general system
\eqref{chemphys:model2}.
We start with a~system with an additive periodic signal, described by the
following equation:

\begin{equation}\label{chemphys:model3}
\dot a = -(k_a + \kappa\xi(t))a 
+ \lambda c\,\xi(t) + \lambda\sqrt{1-c^2}\,\eta(t) + A\cos(\Omega t + \phi).
\end{equation}

\noindent The equation \eqref{chemphys:model3} has a formal solution

\begin{eqnarray}\label{chemphys:solution}
a(t) &=& e^{-k_at}\exp\left[-\kappa\int\limits_0^t\xi(t^\prime)dt^\prime\right]
a_0
\nonumber\\
&&{}
+\lambda  \int\limits_0^t e^{-k_a(t-t^\prime)}
\exp\left[-\kappa\int\limits_{t^\prime}^t\xi(t^{\prime\prime})dt^{\prime\prime}\right]
\left(c\,\xi(t^\prime)+\sqrt{1-c^2}\,\eta(t^\prime)\right)dt^\prime
\nonumber\\
&&{}
+A\int\limits_0^t e^{-k_a(t-t^\prime)}
\exp\left[-\kappa\int\limits_{t^\prime}^t\xi(t^{\prime\prime})dt^{\prime\prime}\right]
\cos(\Omega t^\prime+\phi)\,dt^\prime\,,
\end{eqnarray}

\noindent where $a_0$ is the initial value of the process. For simplicity, we assume
in what follows that $a_0\equiv0$.

For the expectation value of the process we obtain

\begin{eqnarray}\label{chemphys:expectation}
\left\langle a(t)\right\rangle &=& 
\frac{A\cos(\Omega t+\phi +\widetilde\phi)}{\sqrt{(k_a-\frac{1}{2}\kappa^2)^2+\Omega^2}}
-\frac{\lambda\kappa c}{2(k_a-\frac{1}{2}\kappa^2)}
\nonumber\\
&+&
e^{-(k_a-\frac{1}{2}\kappa^2)t}
\left(
\frac{\lambda\kappa c}{2(k_a-\frac{1}{2}\kappa^2)}
-\frac{A\cos(\phi +\widetilde\phi)}{\sqrt{(k_a-\frac{1}{2}\kappa^2)^2+\Omega^2}}
\right),
\end{eqnarray}

\noindent where $\tan\widetilde\phi = -\Omega/(k_a-\frac{1}{2}\kappa^2)$.

As we can see, only the part of the additive noise that is correlated with the
multiplicative noise contributes to the expectation value \eqref{chemphys:expectation}.
There is a~maximal admissible level of the multiplicative noise,
$\kappa_{\mathrm{max}} = \sqrt{2k_a}$, beyond which the process diverges. In the
asymptotic regime, $t\to\infty$, the system displays stationary
oscillations. The amplitude of these oscillations grows monotonically as the 
multiplicative noise increases from zero to the maximal level. This is a~manifestation
of a constructive role of the noise and it is essentially the same as in the
case without the additive noise \cite{JSP}. There is one important difference, though:
Without the correlated additive noise, $\left\langle a(t)\right\rangle$ oscillates
around zero, and with this noise present, it oscillates around a value that depends
on the correlation level and strengths of both the multiplicative and additive noises.
If we now calculate the amplitude of the oscillations relative to this flat background,

\begin{equation}\label{chemphys:relative}
\frac{\mathrm{oscillations\ amplitude}}{\mathrm{background}}
=\frac{A}{\lambda\kappa|c|}\cdot
\frac{2(k_a-\frac{1}{2}\kappa^2)}{\sqrt{(k_a-\frac{1}{2}\kappa^2)^2+\Omega^2}}\,,
\end{equation}

\noindent we can see that the relative amplitude formally diverges for 
$\lambda\kappa c\to0$ (there is no threshold and the system can transfer the
periodic signal even in the absence of the noise) and monotonically decreases
to zero as the level of the multiplicative noise increases towards its maximal value. 
In other words, if the strength of the multiplicative noise is too large, the periodic
stimulation gets drowned in the flat background. In this respect correlations between 
the multiplicative and additive noises act \textit{destructively} on the system.

If we take an average not only over realizations of the noises but also over
the phase of the incoming signal, all oscillations vanish:

\begin{equation}\label{chemphys:doubleaverage2}
\left\langle\left\langle a(t)\right\rangle\right\rangle
\mathop{\longrightarrow}\limits_{t\to\infty}
a_\infty=
-\frac{\lambda\kappa c}{2(k_a-\frac{1}{2}\kappa^2)}\,.
\end{equation}

We now turn to calculating the correlations for the process $a(t)$. The full
expressions are rather long, and therefore we present the results in the
asymptotic ($t\to\infty$) regime only and after averaging on both the realizations
of the noises and the initial phase:

\begin{gather}
\left\langle\left\langle 
a\left(t-{\textstyle\frac{1}{2}}\tau\right)a\left(t+{\textstyle\frac{1}{2}}\tau\right)
\right\rangle\right\rangle
\;\mathop{\longrightarrow}\limits_{t\to\infty}\;
\left(\frac{\lambda\kappa c}{2(k_a-\frac{1}{2}\kappa^2)}\right)^2 +
\frac{A^2\cos\Omega\tau}{2((k_a-\frac{1}{2}\kappa^2)^2+\Omega^2)}
+{}
\nonumber\\
\left(
\frac{A^2\kappa^2}
{4(k_a{-}\kappa^2)\left((k_a{-}\frac{1}{2}\kappa^2)^2+\Omega^2\right)}
{+}\frac{\lambda^2}{2(k_a{-}\kappa^2)}
+\frac{(\lambda\kappa c)^2(k_a-\frac{1}{4}\kappa^2)}{2(k_a-\kappa^2)(k_a{-}\frac{1}{2}\kappa^2)^2}
\right)\!
e^{-(k_a-\frac{1}{2}\kappa^2)\tau}
\label{chemphys:correlation}
\end{gather}

We can see that the correlation function \eqref{chemphys:correlation} exists
only if the multiplicative noise level satisfies 
$\kappa<\sqrt{k_a}=\kappa_{\mathrm{max}}/\sqrt{2}$. For the multiplicative
noise levels in the range 
$\kappa_{\mathrm{max}}/\sqrt{2}\le\kappa<\kappa_{\mathrm{max}}$, the first
moment of the process $a(t)$ converges, but the second does not.

The first term in \eqref{chemphys:correlation} corresponds to the constant
shift introduced by this part of the additive noise that is correlated to the
multiplicative one; it is equal to $a_\infty^2$. The second describes the
oscillations in the correlation function introduced by the external periodic
forcing in \eqref{chemphys:model2}. The remaining terms describe the diffusive
background: First of them is present even in the absence of the additive
noise. The second contains contributions from both the correlated and uncorrelated
parts of the additive noise, and only the correlated part contributes to the
remaining one. Note that the sign of this term is always positive: the presence
of correlations raises the diffusive background. The correlations act destructively
on the system with a additive signal.

The correlation function \eqref{chemphys:correlation}, leads to, via 
Eq.~\eqref{chemphys:appB4}, to the following power spectrum:

\begin{equation}\label{chemphys:power}
P(\omega) =
P_0\,\delta(\omega) +
P_{\mathrm{signal}}\left(\delta(\omega-\Omega)+\delta(\omega+\Omega)\right)
+ P_{\mathrm{back}}(\omega)\,,
\end{equation}

\noindent where

\begin{subequations}
\begin{equation}
P_0 = \left(\frac{\lambda\kappa c}{2(k_a-\frac{1}{2}\kappa^2)}\right)^2
\end{equation}

\noindent is the power density introduced at zero frequency by the constant
shift,

\begin{equation}\label{chemphys:psignal}
P_{\mathrm{signal}} = 
\frac{A^2}{(k_a-\frac{1}{2}\kappa^2)^2+\Omega^2}
\end{equation}

\noindent is the power density associated with the oscillatory 
term, and 

\begin{gather}
P_{\mathrm{back}}(\omega) ={}
\nonumber\\
\frac{A^2\kappa^2(k_a-\frac{1}{2}\kappa^2)^2 + 
\lambda^2\left[2(k_a-\frac{1}{2}\kappa^2)^2 + c^2\kappa^2(k_a-\frac{1}{4}\kappa^2)\right]
\left[(k_a-\frac{1}{2}\kappa^2)^2+\Omega^2\right]}
{4(k_a-\kappa^2)(k_a-\frac{1}{2}\kappa^2)[(k_a-\frac{1}{2}\kappa^2)^2+\Omega^2]
[(k_a-\frac{1}{2}\kappa^2)^2+\omega^2]}
\label{chemphys:pback}
\end{gather}
\end{subequations}

\noindent is the power density associated with the diffusive 
background. We can now calculate the signal-to-noise ratio (SNR) as the ratio between
the power densities transferred by the signal and by the diffusive background at the
frequency of the signal:

\begin{gather}
\mathrm{SNR}(\kappa^2,Z,c)=
\frac{P_{\mathrm{signal}}}
{P_{\mathrm{back}}(\omega=\pm\Omega)}
={}
\nonumber\\
\frac{(k_a-\kappa^2)(k_a-\frac{1}{2}\kappa^2)[(k_a-\frac{1}{2}\kappa^2)^2+\Omega^2]}
{\kappa^2(k_a-\frac{1}{2}\kappa^2)^2 + 
Z\left[(k_a-\frac{1}{2}\kappa^2)^2+\Omega^2\right]
\left[2(k_a-\frac{1}{2}\kappa^2)^2 + c^2\kappa^2(k_a-\frac{1}{4}\kappa^2)\right]}\,,
\label{chemphys:SNR}
\end{gather}

\noindent where $Z=(\lambda/A)^2$ measures the relative strength of the additive noise 
and the signal. Regardless of the values of other parameters, SNR drops to zero for
$\kappa^2=k_a$, meaning that the total power density comes from the noise and the 
signal is totally undistinguishable from it. For $Z=0$
\eqref{chemphys:SNR} formally diverges for $\kappa\to0$ (recall that since there
is no threshold, the system transfers the signal even for a vanishing multiplicative noise).
SNR remains finite for $Z>0$ and
since the second term in the denominator of \eqref{chemphys:SNR} is nonnegative,
$\mathrm{SNR}(\kappa^2,Z,c)\leqslant\mathrm{SNR}(\kappa^2,0,\cdot)$.
If the above inequality is strong, the presence of the additive noise acts
destructively on the system's capabilities to transfer the signal.
In any case, SNR decreases monotonically as the multiplicative noise increases: 
there is no stochastic resonance.

For weak signals ($Z\geqslant1$) the SNR scales as $Z^{-1}$, which is readily seen in
\eqref{chemphys:SNR}.
Furthermore, $\forall\,Z>0, \kappa: \mathrm{SNR}(\kappa^2,Z,c\not=0)<
\mathrm{SNR}(\kappa^2,Z,0)$: If the additive and multiplicative noises are correlated,
the system's capabilities to transfers the periodic signal are slightly worse than
those of the system with uncorrelated noises.
Numerical simulations show that the system can transfer signals as weak as $Z\lessim100$,
but for such a weak signal the difference between the $c=0$ and $|c|=1$ cases is scarcely
noticeable. 

Finally, we may ask what is the relative ``height'' of the $\delta$-peak in the 
spectrum associated with the signal to the $\delta$-peak associated with what 
manifests itself as a~constant forcing and results from the presence of the 
correlations. Not surprisingly,

\begin{equation}\label{chemphys:peaks}
\frac{P_{\mathrm{signal}}}{P_0} =
\frac{A^2}{(\lambda\kappa c)^2}\cdot
\frac{4(k_a-\frac{1}{2}\kappa^2)^2}{(k_a-\frac{1}{2}\kappa^2)^2+\Omega^2}\,,
\end{equation}

\noindent which once again shows that in the presence of the correlations between the
multiplicative and additive noises, the signal is drowned by the correlated noise
as the multiplicative noise level increases. Note that the ratio \eqref{chemphys:peaks}
diverges if $\lambda\kappa c=0$; this reflects the fact that if any of the noises
vanishes, or if the noises are uncorrelated, there is no constant shift in
the output, cf.~Eq.~\eqref{chemphys:expectation} above.

\section{A multiplicative signal}\label{chemphys:section-multiplicative}

We now consider a model described by the following equation:

\begin{equation}\label{chemphys:parametrical}
\dot a = -(k_a + \kappa\xi(t) + A\cos(\Omega t+\phi))a + 
\lambda c\,\xi(t) + \lambda\sqrt{1-c^2}\,\eta(t)\,,
\end{equation}

\noindent where all quantities have meanings and properties of the corresponding
quantities in Eq.~\eqref{chemphys:model3} above. The only difference is that 
the signal is now coupled parametrically to the transmitter. We have discussed
this model in Ref.~\cite{PRE2001}, but only for the uncorrelated case ($c=0$),
where we have shown that because of the presence of the additive noise, the
system responds, in terms of its autocorrelation function, to the signal even
though the average realization of $a(t)$ goes to zero.

\subsection{Analytical results}

A formal solution to Eq.~\eqref{chemphys:parametrical} reads

\begin{eqnarray}\label{chemphys:parametrical-solution}
a(t) &=&
\lambda\int\limits_0^t e^{-k_a(t-t^\prime)}
\exp\left[-\kappa\int\limits_{t^\prime}^t\xi(t^{\prime\prime})dt^{\prime\prime}
-A\int\limits_{t^\prime}^t\cos(\Omega t^{\prime\prime} +\phi)dt^{\prime\prime}
\right]
\times\nonumber\\
&&\quad
\left(c\,\xi(t^\prime)+\sqrt{1-c^2}\,\eta(t^\prime)\right)dt^\prime\,,
\end{eqnarray}

\noindent where we have assumed that the initial value $a(0)=0$ for simplicity.
As in the previous Section, we will use this formal solution in calculating 
the expectation value of the 
process $a(t)$ and its autocorrelation function. The calculations run similarly 
to those for the additive noise and make use of the expectation values evaluated in 
the Appendix. For the expectation value we obtain

\begin{eqnarray}\label{chemphys:expect-multi1}
\left\langle a(t) \right\rangle
&=&-\frac{1}{2}\lambda\kappa c
\int\limits_0^t e^{-(k_a-\frac{1}{2}\kappa^2)(t-t^\prime)}
\times{}
\nonumber\\
&&\quad
\exp\left[-\frac{2A}{\Omega}\sin\frac{1}{2}\Omega(t-t^\prime)
\cos\left(\phi+\frac{1}{2}\Omega(t+t^\prime)\right)\right] dt^\prime\,.
\end{eqnarray}

If the signal is weak, $\vert A/\Omega\vert\ll1$, we may expand the expectation 
value \eqref{chemphys:expect-multi1}, not averaged over the initial phase,
to obtain

\begin{subequations}\label{chemphys:expand1}
\begin{eqnarray}
\left\langle a(t)\right\rangle 
&\mathop{\longrightarrow}\limits_{t\to\infty}&
-\frac{c\lambda\kappa}{2(k_a-\frac{1}{2}\kappa^2)}
\left(1+\frac{A}{\Omega}\,
\frac{\cos(\Omega t +\phi +\widetilde\phi)}{\sqrt{1+\tan^2\widetilde\phi}}
\right),
\\
\tan\widetilde\phi &=& \frac{k_a-\frac{1}{2}\kappa^2}{\Omega}\,.
\end{eqnarray}
\end{subequations}

\noindent We can see that in the limit of weak signals, the transmitter
asymptotically displays a constant shift and oscillations. The amplitude
of these oscillations relative to the constant background grows steadily
as the multiplicative noise level increases towards $\kappa_{\text{max}}$:
The oscillations induced by a parametrically coupled signal
are \textit{not} drowned in the noisy background. The presence of these oscillations
is clearly a constructive effect of the correlations as in the uncorrelated
case, $c=0$, $\left\langle a(t)\right\rangle\equiv0$.

\noindent If we further average Eq.~\eqref{chemphys:expect-multi1} over the 
phase of the incoming signal, we obtain

\begin{eqnarray}\label{chemphys:expect-multi2}
\left\langle\left\langle a(t)\right\rangle\right\rangle &=& 
-\frac{1}{2}\lambda\kappa c
\int\limits_0^t dt^\prime\, e^{-(k_a-\frac{1}{2}\kappa^2)(t-t^\prime)}
\times{}
\nonumber\\
&&\quad
\frac{1}{2\pi}\int\limits_0^{2\pi} d\phi
\exp\left[-\frac{2A}{\Omega}\sin\frac{1}{2}\Omega(t-t^\prime)
\cos\left(\phi+\frac{1}{2}\Omega(t+t^\prime)\right)\right]
\nonumber\\
&=&
-\frac{1}{2}\lambda\kappa c
\int\limits_0^t dt^\prime\, e^{-(k_a-\frac{1}{2}\kappa^2)t^\prime}
\frac{1}{2\pi}\int\limits_0^{2\pi} d\phi
\exp\left[-\frac{2A}{\Omega}\sin\frac{1}{2}\Omega t^\prime
\cos\phi\right],
\end{eqnarray}

\noindent since the integration over $\phi$ runs over the entire period
of the integrand and therefore the value of the inner integral cannot
depend on $\frac{1}{2}\Omega(t+t^\prime)$. An integral representation
of the modified Bessel function

\begin{equation}\label{chemphys:bessel}
I_0(z) = \sum\limits_{n=0}^\infty \frac{(z/2)^{2n}}{(n!)^2}
=\frac{1}{2\pi}\int\limits_0^{2\pi} \exp\left(z\cos\phi\right)d\phi
\end{equation}

\noindent can immediately be recognized \cite{Abramowitz}:

\begin{equation}\label{chemphys:expect-multi3}
\left\langle\left\langle a(t)\right\rangle\right\rangle
=
-\frac{1}{2}c\lambda\kappa \int\limits_0^t
e^{-(k_a-\frac{1}{2}\kappa^2)t^\prime}
I_0\left(\frac{2A}{\Omega}\sin\frac{1}{2}\Omega t^\prime\right) dt^\prime\,.
\end{equation}

\noindent Furthermore, as

\begin{gather}
0\leqslant 
\int\limits_t^\infty e^{-(k_a-\frac{1}{2}\kappa^2)t^\prime}
I_0\left(\frac{2A}{\Omega}\sin\frac{1}{2}\Omega t^\prime\right) dt^\prime
\leqslant
\int\limits_t^\infty e^{-(k_a-\frac{1}{2}\kappa^2)t^\prime}
I_0\left(\frac{2A}{\Omega}\right) dt^\prime
\nonumber\\
{}=
\frac{1}{k_a-\frac{1}{2}\kappa^2}\,I_0\left(\frac{2A}{\Omega}\right)
e^{-(k_a-\frac{1}{2}\kappa^2)t}
\mathop{\longrightarrow}\limits_{t\to\infty} 0\,,
\label{chemphys:szacunek}
\end{gather}

\noindent we may conclude that

\begin{equation}\label{chemphys:expect-multi4}
\left\langle\left\langle a(t)\right\rangle\right\rangle
\mathop{\longrightarrow}\limits_{t\to\infty}
-\frac{1}{2}\lambda\kappa c
\int\limits_0^\infty e^{-(k_a-\frac{1}{2}\kappa^2)t^\prime}
I_0\left(\frac{2A}{\Omega}\sin\frac{1}{2}\Omega t^\prime\right) dt^\prime\,
\end{equation}

\noindent provided that $\kappa<\kappa_{\text{max}}=\sqrt{2k_a}$; otherwise
the process is divergent, which is readily seen directly from 
Eq.~\eqref{chemphys:expect-multi1}. Observe that if $A\geqslant 
k_a-\frac{1}{2}\kappa^2$, the expectation value
\eqref{chemphys:expect-multi4} formally diverges in the limit $\Omega\to0$,
even though realizations for some particular values of the initial phase $\phi$
may be convergent.

For the correlation function we obtain

\begin{equation}\label{chemphys:multi-correlation}
\left\langle\left\langle 
a\left(t-{\textstyle\frac{1}{2}}\tau\right)a\left(t+{\textstyle\frac{1}{2}}\tau\right)
\right\rangle\right\rangle
=
J_1+ J_2 + J_3\,,
\end{equation}

\noindent where

\begin{subequations}\label{chemphys:multi-correlationJ}
\begin{eqnarray}
J_1 &=& \lambda^2 e^{-(k_a-\frac{1}{2}\kappa^2)\tau} 
\int\limits_0^t ds\, e^{-2(k_a-\kappa^2)s} I_0\left(\frac{A}{\Omega}Z_1\right),
\\
J_2 &=& {\textstyle\frac{3}{2}}(\lambda\kappa c)^2
e^{-(k_a-\frac{1}{2}\kappa^2)\tau}
\int\limits_0^{t-\frac{1}{2}\tau} ds\, e^{-2(k_a-\kappa^2)s}
\int\limits_0^{t-\frac{1}{2}\tau-s} du\, e^{-(k_a-\frac{1}{2}\kappa^2)u}
I_0\left(\frac{A}{\Omega}Z_2\right),
\nonumber\\\\
J_3 &=& {\textstyle\frac{1}{4}}(\lambda\kappa c)^2
\int\limits_0^{t-\frac{1}{2}\tau} ds\, e^{-(k_a-\frac{1}{2}\kappa^2)s}
\int\limits_0^\tau du\, e^{-(k_a-\frac{1}{2}\kappa^2)u}
I_0\left(\frac{A}{\Omega}Z_3\right),
\end{eqnarray}
\end{subequations}

\noindent and $Z_i$ are certain functions involving trigonometric functions of
$\tau$ and the integration variables, but \textit{not} of~$t$.
We can see that this correlation function converges for $\kappa<\sqrt{k_a}
=\kappa_{\text{max}}/\sqrt{2}$. A~reasoning similar to that used in 
\eqref{chemphys:szacunek}
shows that the correlation function becomes stationary in the limit $t\to\infty$.
In the limit $\Omega\to0$, the correlation function 
\eqref{chemphys:multi-correlation} formally diverges if $A>2(k_a-\kappa_2)$.
Note also that the square of the strength of the additive noise $\lambda$ multiplies all
terms in \eqref{chemphys:multi-correlationJ}.
The situation is slightly paradoxical: without the
additive noise, almost all realizations of the process $a(t)$ eventually die out,
yet as in the output process the power attributed to ``signal'' and to ``noise''
scale by $\lambda^2$, the SNR does not depend on the additive noise level.

\subsection{Numerical results}

\begin{figure}
\begin{center}
\includegraphics[scale=0.77]{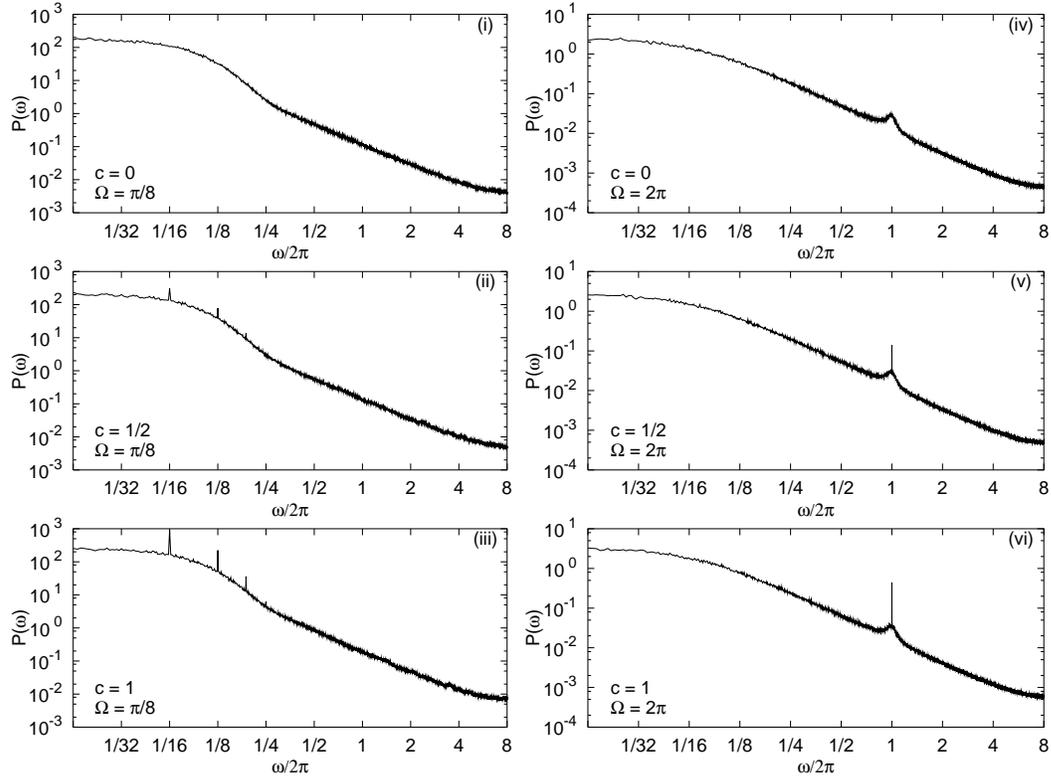}
\end{center}
\caption{Numerical power spectra of the process \eqref{chemphys:parametrical}
for various input signal frequencies and correlations between the noises.
The input signal frequency equals $\Omega=\pi/8$ (panels (i) through (iii))
and $\Omega=2\pi$ (panels (iv) through (vi)). The multiplicative and
additive noises are uncorrelated ($c=0$) on panels (i), (iv), partially
correlated ($c=1/2$) on panels (ii), (v), and fully correlated ($c=1$)
on panels (iii), (vi). Other parameters, common for all panels, are $k_a=1/2$,
$\kappa =\sqrt{k_a}/2$, $\lambda=1/4$, and $A=1$.}
\label{chemphys:fig1}
\end{figure}

Because of the analytical structure of Eq.~\eqref{chemphys:multi-correlationJ}, any
further analysis of this correlation function should be performed numerically.
Instead of numerically calculating the Fourier transform of 
Eq.~\eqref{chemphys:multi-correlation},
we have directly integrated Eq.~\eqref{chemphys:parametrical} by the Heun method
\cite{Manella} with a time step $h=1/256$, calculated the power spectrum for each
realization and averaged over 512 realizations of the noises and the initial phase.
We have used Marsaglia algorithm \cite{Marsaglia} to generate
the GWNs and the famous Mersenne Twister \cite{Mersenne} as the underlying
uniform generator.
Selected results are presented on Fig.~\ref{chemphys:fig1}. As we can see, 
if the external frequency is small and the two noises are uncorrelated, the
process $a(t)$ does not display any oscillations, cf.~panel~(i). If the noises become
correlated, distinct sharp peaks corresponding to the driving frequency and its 
higher harmonics appear, cf.~panels~(ii) and~(iii). For a large driving frequency, 
in the uncorrelated case there is a broad peak
centered at the driving frequency, cf.~\ panel~(iv). If the noises get correlated, 
a sharp line appears, superimposed on the broad peak, cf.\ panels~(v) and~(vi).

If there are no correlations, oscillations introduced by a parametrically
coupled signal are, by their very nature, damped with an effective damping
constant $k_a-\frac{1}{2}\kappa^2$. If the driving frequency is very small,
with $\Omega^{-1}\gg(k_a-\frac{1}{2}\kappa^2)^{-1}$, the transmitter is dominated
by the damping term and the resulting power spectrum is non-oscillatory
(Fig.~\ref{chemphys:fig1}, panel~(i) above). For larger frequencies a broad peak,
characteristic for damped oscillations, appears in the power spectrum
(Fig.~\ref{chemphys:fig1}, panel~(iv)). The presence 
of correlations between the additive and multiplicative noises changes the
situation. Sharp lines, characteristic for undamped oscillations, appear in the
power spectrum for both small and large driving frequencies. We can see that
the correlations have a strong \textit{constructive} effect on the transmitted
signal in case of a~multiplicative coupling.

The presence of higher harmonics of the driving frequency on panels (ii), (iii)
reflects a nonlinear nature of the multiplicative coupling between the
transmitter and the external signal.

\subsection{The origin of the constructive role of correlations}

To understand, at least qualitatively, why correlations between 
the additive and multiplicative noises reinforce the parametrically
coupled signal, observe
that Eq.~\eqref{chemphys:parametrical} is formally equivalent to

\begin{eqnarray}\label{chemphys:multi-modified}
\dot a &=& -\left(k_a + \kappa\xi(t) + A\cos(\Omega t+\phi)\right)
\left(a-\frac{c\lambda}{\kappa}\right)
\nonumber\\
&&{}
-\frac{c\lambda}{\kappa} k_a -\frac{c\lambda}{\kappa}A\cos(\Omega t+\phi)
+\lambda\sqrt{1-c^2}\,\eta(t)\,.
\end{eqnarray}

\noindent If we now introduce 

\begin{equation}\label{chemphys:transformation}
\widetilde{a} = a -c\lambda/\kappa\,,
\end{equation}

\noindent we obtain

\begin{equation}\label{chemphys:multi-atilde}
\dot{\widetilde{a}} = -\left(k_a + \kappa\xi(t) + A\cos(\Omega t+\phi)\right)\widetilde{a}
-\frac{c\lambda}{\kappa} k_a -\frac{c\lambda}{\kappa}A\cos(\Omega t+\phi)
+\lambda\sqrt{1-c^2}\,\eta(t)\,.
\end{equation}

\noindent The correlated part of the noise has been formally removed from
Eq.~\eqref{chemphys:multi-atilde}.
We can now see that the presence of correlations effectively amounts 
to (i) lowering of the amplitude of the additive noise, 
(ii) introducing a constant additive forcing, and (iii) additively coupling 
to the transmitter a~periodic signal with exactly the same phase and frequency 
as the multiplicative one. The amplitude of the additional signal 
is proportional to the product of the correlation coefficient and the strength
of the additive noise. As we already know from Section~\ref{chemphys:section-additive},
an additive periodic signal introduces undamped oscillations into the correlation
function which, in turn, introduce very sharp lines into the power spectrum.
Such lines are clearly visible on Fig.~\ref{chemphys:fig1}.

Note that a transformation analogous to \eqref{chemphys:transformation},
when applied to the system with a~purely additive signal
\eqref{chemphys:model3}, does not affect the signal.
This observation, perhaps, clarifies why the presence of correlations
in the additive case does not lead to any constructive effects.

\section{Discussion}\label{chemphys:section-discussion}

In this paper we have discussed the effects of correlations between additive
and multiplicative noises in a model of a linear transmitter \eqref{chemphys:model2}.
We have shown that the effects of correlations is different depending on whether
the signal is coupled additively or multiplicatively. In the former case, 
the correlations act destructively: The amplitude of
the outgoing signal increases with the level of the multiplicative noise, but
the constant shift introduced by the correlations increases even faster, 
the amplitude of the oscillations
relative to the shift decreases, and the signal is eventually drowned by the
noise. Note that the correlations act here against the noise: The multiplicative
noise enhances the signal, but the correlations diminish it. 
Similar effects can be seen in the power spectrum and the SNR of the outgoing
signal. The fact that correlations
reduce the destructive effects of noise has been reported previously; papers 
quoted in the Introduction provide examples of such systems.
What we have here is an 
example of correlations reducing the constructive effect of noise.

These formal results, and the correlation function \eqref{chemphys:correlation}
in particular, agree with those reported in Ref.~\cite{berdichevsky99},
where the Gaussian white noise has been obtained as a limiting case of
a dichotomous, colored noise. This should come as no surprise: There is 
a generally held belief (see e.g. \cite{chris}) that in a well defined limit
the two, in principle very different, kinds of noise have identical effect
on the dynamics of a system. A comparison of our present results with those
of Ref.~\cite{berdichevsky99} provides yet another confirmation of this statement.

The situation is entirely different for the case of a multiplicative signal.
We have shown that for a weak incoming signal \textit{not} averaged
over the phase, the outgoing signal displays oscillations that do not vanish
as the level of the multiplicative noise increases. Numerical evidence suggests
that without the correlations, the power spectrum averaged over both realizations
of the noise and the initial phase does not display any signs of oscillations
for low frequencies and only damped oscillations for larger frequencies. 
This is easy to understand: A multiplicative signal is, by its very nature, damped,
and if the wavelength of the signal is large, the signal does not recover
before if gets completely washed away by the damping. The presence of correlations
changes this picture altogether: The power spectrum displays distinct, sharp
lines for all frequencies of the input signal. We have showed analytically that 
correlations between
the multiplicative and additive noises are effectively equivalent to adding
an extra signal, with exactly the same phase and frequency as the multiplicative
one, to the system. The amplitude of this extra signal is proportional to the
correlation coefficient and it is this signal that produces the sharp lines
($\delta$-peaks) in the power spectrum.

We have assumed in Section~\ref{chemphys:section-model} that the two noises present
in the system together form a two-dimensional Gaussian process. On a formal level,
this property is reflected in the relation between the amplitudes of the correlated
and uncorrelated parts of the additive noise. If we relax this assumption, instead
of \eqref{chemphys:model2} we obtain

\begin{equation}\label{chemphys:model2-relaxed}
\dot a = -(k_a + \kappa\xi(t)+A_{\text{m}}\cos(\Omega t+\phi))a + 
\lambda_\xi\xi(t) + \lambda_\eta\eta(t)
+A_{\text{a}}\cos(\Omega t+\phi),
\end{equation}

\noindent where no special dependence between $\lambda_\xi$ and $\lambda_\eta$
is assumed. The
results reported in this paper can now be recovered if $c^2$ is replaced by
$\lambda_\xi^2/(\lambda_\xi^2+\lambda_\eta^2)$ and $Z=(\lambda/A_{\text{a}})^2$ is replaced
by $(\lambda_\xi^2+\lambda_\eta^2)/A_{\text{a}}^2$. Thus, results reported
do not depend on the assumption that the \textit{joint} distribution of
the noises is Gaussian. They do depend, however, on the assumption that each 
individual noise is Gaussian and point-correlated in time.

I acknowledge a very helpful discussion with Prof.\ Peter H\"anggi that
helped me to clarify certain ambiguities present in a draft version of this paper.

\appendix

\section{Appendix: The expectation values}

While calculating the expectation value of $a(t)$ and the correlation function of
this process, one encounters several integrals that involve the expectation
value of the exponential of an integral of the multiplicative noise $\xi(t)$.
Specifically,

\begin{equation}\label{chemphys:exp}
\left\langle
\exp\left[\int\limits_0^T \varphi(t^\prime)\xi(t^\prime)\,dt^\prime\right]
\right\rangle 
=
\exp\left[\frac{1}{2}\int\limits_0^T [\varphi(t^\prime)]^2\,dt^\prime\right],
\end{equation}

\noindent where $T$ is a certain time and $\varphi(t)$ is a certain function,
not necessarily continuous (cf.~\cite{JSP,Kubo}). Similarly,

\begin{subequations}\label{chemphys:expeta}
\begin{eqnarray}
\left\langle\eta(t_1)
\exp\left[\int\limits_0^T \varphi(t^\prime)\xi(t^\prime)dt^\prime\right]
\right\rangle 
&=&0\,,
\\
\left\langle\eta(t_1)\eta(t_2)
\exp\left[\int\limits_0^T \varphi(t^\prime)\xi(t^\prime)dt^\prime\right]
\right\rangle 
&=&
\delta(t_1{-}t_2)
\exp\left[\frac{1}{2}\int\limits_0^T [\varphi(t^\prime)]^2dt^\prime\right]
\end{eqnarray}
\end{subequations}

\noindent because $\eta(t)$, $\xi(t)$ are independent GWNs. The other two 
expectation values are more challenging. We start with

\begin{equation}\label{chemphys:expxi1}
C_1=
\left\langle\xi(t_1)
\exp\left[\int\limits_0^T \varphi(t^\prime)\xi(t^\prime)\,dt^\prime\right]
\right\rangle.
\end{equation}

\noindent We expand the exponential function in a Taylor series,
interchange the orders of summations, and obtain

\begin{equation}\label{chemphys:expxi2}
C_1=
\sum\limits_{n=0}^\infty
\frac{1}{n!}\int\limits_0^T\,dt_{\mu_1}\dots\int\limits_0^T\,dt_{\mu_n}
\varphi(t_{\mu_1})\dots\varphi(t_{\mu_n})
\left\langle\xi(t_{\mu_1})\dots\xi(t_{\mu_n})\xi(t_1)\right\rangle.
\end{equation}

\noindent The expectation value on the right-hand side of \eqref{chemphys:expxi2}
involves $n{+}1$ terms. Therefore, only the terms with $n=2m{+}1$ will give a nonzero
contribution to the sum. The expectation value thus involves $2(m{+}1)$ noise terms.
It factors out to a~product of $m{+}1$ two-point correlations and there are $(2m{+}1)!!$
ways to choose the pairs. In every possible choice, one of the pairs involves $t_1$
and a~certain $t_\mu$, and the remaining $m$ pairs involve two $t_\mu$'s. Thus

\begin{eqnarray}\label{chemphys:expxi3}
 C_1 &=& \sum\limits_{m=0}^\infty \frac{(2m+1)!!}{(2m+1)!}
\left(
\int\limits_0^T \varphi(t_\mu)\left\langle\xi(t_\mu)\xi(t_1)\right\rangle\,dt_\mu
\right) \times{}
\nonumber\\
&&
\hphantom{\sum\limits_{m=0}^\infty \frac{(2m+1)!!}{(2m+1)!}}
\left(\int\limits_0^T dt_{\mu_1} \int\limits_0^T dt_{\mu_2}
\varphi(t_{\mu_1})\varphi(t_{\mu_2})
\left\langle\xi(t_{\mu_1})\xi(t_{\mu_2})\right\rangle\right)^m
\nonumber\\
&=&
\sum\limits_{m=0}^\infty \frac{1}{2^m m!}\, \varphi(t_1)
\left(\int\limits_0^T \left[\varphi(t_\mu)\right]^2\,dt_\mu\right)^m
\nonumber\\
&=&
\varphi(t_1)\exp\left[
\frac{1}{2}\int\limits_0^T \left[\varphi(t_\mu)\right]^2\,dt_\mu
\right].
\end{eqnarray}

The last remaining expectation value involves two noise terms multiplying
the exponential:

\begin{eqnarray}\label{chemphys:expxixi1}
C_2 &=&
\left\langle\exp\left[
\int\limits_0^T \varphi(t^\prime)\xi(t^\prime)\,dt^\prime
\right]\xi(t_1)\xi(t_2)\right\rangle
\nonumber\\
&=&
\sum\limits_{n=0}^\infty
\frac{1}{n!}\int\limits_0^T\,dt_{\mu_1}\dots\int\limits_0^T\,dt_{\mu_n}
\varphi(t_{\mu_1})\dots\varphi(t_{\mu_n})
\left\langle\xi(t_{\mu_1})\dots\xi(t_{\mu_n})\xi(t_1)\xi(t_2)\right\rangle,
\nonumber\\
\end{eqnarray}

\noindent where, as above, we have expanded the exponential into the power series.
Only the terms with $n=2m$ will give a nonzero contribution to the sum. The
expectation value of the product of noise terms factors out to a product of 
two-point correlations. Again, there are $(2m{+}1)!!$ ways to choose the pairs.
In $(2m{-}1)!!$ cases the noises at $t_1$, $t_2$ will be coupled with each
other, and in the remaining cases they will be coupled with different $t_\mu$s.
Therefore

\begin{eqnarray}\label{chemphys:expxixi2}
C_2 &=&
\sum\limits_{m=0}^\infty \frac{(2m-1)!!}{(2m)!} \left\langle\xi(t_1)\xi(t_2)\right\rangle
\times{}
\nonumber\\
&&
\hphantom{\sum\limits_{m=0}^\infty}
\left(\int\limits_0^T dt_{\mu_1} \int\limits_0^T dt_{\mu_2}
\varphi(t_{\mu_1})\varphi(t_{\mu_2})
\left\langle\xi(t_{\mu_1})\xi(t_{\mu_2})\right\rangle\right)^m
\nonumber\\
&+&
\sum\limits_{m=1}^\infty \frac{(2m+1)!!-(2m-1)!!}{(2m)!}
\times{}
\nonumber\\
&&
\hphantom{\sum\limits_{m=1}^\infty}
\left(
\int\limits_0^T \varphi(t_\mu)\left\langle\xi(t_\mu)\xi(t_1)\right\rangle\,dt_\mu
\right)
\left(
\int\limits_0^T \varphi(t_\nu)\left\langle\xi(t_\nu)\xi(t_2)\right\rangle\,dt_\nu
\right)
\times{}
\nonumber\\
&&
\hphantom{\sum\limits_{m=1}^\infty}
\left(\int\limits_0^T dt_{\mu_1} \int\limits_0^T dt_{\mu_2}
\varphi(t_{\mu_1})\varphi(t_{\mu_2})
\left\langle\xi(t_{\mu_1})\xi(t_{\mu_2})\right\rangle\right)^{m-1}
\nonumber\\
&=&
\delta(t_1-t_2)
\sum\limits_{m=0}^\infty \frac{1}{2^m m!}
\left(\int\limits_0^T \left[\varphi(t_\mu)\right]^2\,dt_\mu\right)^m
\nonumber\\
&+&
\varphi(t_1)\varphi(t_2)
\sum\limits_{m=1}^\infty \frac{1}{2^{m-1} (m-1)!}
\left(\int\limits_0^T \left[\varphi(t_\mu)\right]^2\,dt_\mu\right)^{m-1}
\nonumber\\
&=&
\left[\delta(t_1-t_2)+\varphi(t_1)\varphi(t_2)\right]
\exp\left[
\frac{1}{2}\int\limits_0^T \left[\varphi(t_\mu)\right]^2\,dt_\mu
\right].
\end{eqnarray}

This approach can be easily generalized to colored Gaussian noises. Suppose
$\xi(t)$ is a Gaussian noise with zero mean and correlations
of the form

\begin{equation}\label{chemphys:color1}
\left\langle\xi(t)\xi(t^\prime)\right\rangle = h(|t-t^\prime|)\,,
\end{equation}

\noindent where $\int_{-\infty}^\infty h(u)du=1$. Then, instead of 
\eqref{chemphys:expxixi2}, we obtain

\begin{gather}\label{chemphys:color2}
\left\langle\exp\left[
\int\limits_0^T \varphi(t^\prime)\xi(t^\prime)\,dt^\prime
\right]\xi(t_1)\xi(t_2)\right\rangle
={}
\nonumber\\
\left[
h(|t_1-t_2|)
+
\left(\int\limits_0^T \varphi(t_\mu)h(|t_\mu-t_1|)\,dt_\mu\right)
\left(\int\limits_0^T \varphi(t_\nu)h(|t_\nu-t_2|)\,dt_\nu\right)
\right]
\times{}
\nonumber\\
\exp\left[\frac{1}{2}
\int\limits_0^T dt_\mu \int\limits_0^T dt_\nu\,
\varphi(t_\mu)\varphi(t_\nu)h(|t_\mu-t_\nu|)
\right]
\end{gather}

\noindent and similar expression for other expectation values.

As we have said, the function $\varphi$ needs not to be continuous.
In many practical situations, for example in evaluating expressions like

\begin{equation}\label{chemphys:interval}
\left\langle
\exp\left[\int\limits_{t_1}^{t_2} \varphi(t^\prime)\xi(t^\prime)\,dt^\prime\right]
\right\rangle 
\end{equation}

\noindent with $0\leqslant t_1<t_2\leqslant T$, $\varphi(t)$ would be the characteristic
function of an interval:

\begin{equation}\label{chemphys:characteristic}
\varphi(t)=
\begin{cases}
0&t<t_1\\
\frac{1}{2}&t=t_1\\
1&t_1<t<t_2\\
\frac{1}{2}&t=t_2\\
0&t>t_2
\end{cases}
\end{equation}

\noindent The rationale for putting $\frac{1}{2}$ at the endpoints is that

\begin{equation}\label{chemphys:rationale}
\frac{1}{2}=\int\limits_{t_1}^{t_2} \delta(t-t_1)\,dt=
\int\limits_{-\infty}^\infty \varphi(t)\delta(t-t_1)\,dt=
\varphi(t_1)\,.
\end{equation}

\noindent The first equality stems from the fact that in 
$\int_{t_1}^{t_2} \delta(t-t_1)\,dt$
the integrand contains only ``a half of the peak''. By the same token,
for example

\begin{equation}
\left\langle\xi(t_1)\exp\left[-\kappa\int\limits_{t_1}^T\xi(t^\prime)\,dt^\prime
\right]\right\rangle = 
-{\textstyle\frac{1}{2}}\kappa\exp\left({\textstyle\frac{1}{2}}\kappa^2(T-t_1)\right),
\end{equation}

\noindent where $T\geqslant t_1$ and Eq.~\eqref{chemphys:expxi3} has been used.

\end{document}